# Broadband Near-Infrared Compressive Spectral Imaging System with Reflective Structure

YUTONG LI,[1,2] ZHENMING YU,[1,2 *] LIMING CHENG,[1,2] JIAYU DI,[1,2] LIANG LIN,[1,2] JINGYUE MA,[1,2] TONGSHUO ZHANG,[1,2] YUE ZHOU,[1] HAIYING ZHAO,[3] AND KUN XU[1,2]

[1]*State Key Laboratory of Information Photonics and Optical Communications, Beijing University of Posts and Telecommunications, Beijing 100876, China.*
[2] *Xiong' an Aerospace Information Research Institute, Xiong' an New Area 071700, China.*
[3]*School of Artificial Intelligence, Beijing University of Posts and Telecommunications, Beijing 100876, China.*
*\*yuzhenming@bupt.edu.cn*

**Abstract:** Near-infrared (NIR) hyperspectral imaging has become a critical tool in modern analytical science. However, conventional NIR hyperspectral imaging systems face challenges including high cost, bulky instrumentation, and inefficient data collection. In this work, we demonstrate a broadband NIR compressive spectral imaging system that is capable of capturing hyperspectral data covering a broad spectral bandwidth ranging from 700 to 1600 nm. By segmenting wavelengths and designing specialized optical components, our design overcomes hardware spectral limitations to capture broadband data, while the reflective optical structure makes the system compact. This approach provides a novel technical solution for NIR hyperspectral imaging.

## 1. Introduction

Near-infrared (NIR) hyperspectral imaging combines conventional spectroscopy with imaging techniques and enables the capture of three-dimensional data, including spatial and NIR spectral information. NIR has high penetration in the atmosphere and biological tissues, and it is also able to identify material-specific spectral fingerprints. Compared to the visible spectrum, NIR can provide more chemical information and reveal more details. These advantages have led to the extensive application of NIR hyperspectral imaging in areas such as detecting pollutants in water [1], non-destructively analyzing the composition of cultural relics [2,3], and assessing the growth status of vegetation [4,5].

Hyperspectral imaging systems primarily employ tunable filters or scanning techniques. Nonetheless, these methods are fundamentally limited by the Nyquist sampling theorem, resulting in long acquisition times and large data volumes [6,7]. In recent years, snapshot hyperspectral imaging based on compressed sensing has emerged as an innovative solution, with the capability to capture the spatial-spectral data cube in a single exposure [8]. The first CASSI system is a dual-disperser (DD) CASSI [9], which uses two prisms to prevent the spatial shifting and blurred boundaries. In contrast, single-disperser (SD) CASSI utilizes a single prism [10], reducing both the number of optical components and the overall system size. R-CASSI combines the strengths of both systems, achieving a more compact design while maintaining high spatial resolution [11]. Additionally, various studies have focused on developing new modulation methods and hardware architectures for CASSI, intending to improve performance or extend its application areas. Examples include coded-filter-array-based color CASSI systems [12], grating-based CASSI systems [13], and dual-camera CASSI setups [14]. However, despite the advancements in snapshot hyperspectral imaging, most current snapshot hyperspectral imaging systems remain confined to the visible spectrum. The expensive optical detectors increase the cost of infrared two-dimensional acquisition, and the band-pass transmission properties of components limit their ability to cover the entire spectral range. These constraints make snapshot hyperspectral imaging's direct adaptation to the infrared range exceedingly challenging. As a result, there is a growing need for research focused on snapshot hyperspectral imaging in the NIR range.

At present, there are only a few studies focused on infrared hyperspectral imaging. Yang et al. [15] realized efficient compression and reconstruction of mid-infrared spectral data by designing a specialized digital micromirror device (DMD) and employing an optimized reconstruction algorithm. However, their study focused exclusively on the mid-infrared range. Lukáš Klein et al. [16] used a DMD and two single-pixel cameras to acquire hyperspectral data in both the visible and NIR bands. Nevertheless, the single-pixel acquisition is time-consuming and utilizing two cameras results in higher costs. Limited by the high cost of NIR detectors and the stringent demands on spectral dispersion components due to the broad spectral range of the NIR band, new advancements are required to broaden the spectral range and optimize the compactness of existing NIR snapshot imaging systems for practical applications. With the advancement of photonic-chip fabrication technologies, filter-based approaches have emerged as highly compact solutions for hyperspectral imaging, making them suitable for both indoor and outdoor environments, as well as UAV applications. In these designs, a single-chip multispectral filter can play the role of a fixed mask and a disperser in CASSI [17-20]. A recent study has successfully demonstrated spectral acquisition capability spanning 400-1700 nm using this approach [20]. However, due to the immaturity of current manufacturing processes, these methods suffer from low yield rates, limiting their commercial application. In contrast, CASSI offers high flexibility and scalability, and can be tailored for NIR broadband imaging through novel design enhancements.

In this Letter, we develop a compressive spectral imaging system that is capable of capturing broadband NIR hyperspectral data ranging from 700 to 1600 nm in a single exposure. The R-CASSI system captures reflected light using a coded aperture and beam splitter, with a single prism serving as dual dispersers in a highly compact optical path. However, this work was limited to the visible spectrum. Expanding on the R-CASSI foundation, we introduce a wavelength-segmented design and, for the first time, extend CASSI architecture into the NIR spectrum. To overcome spectral efficiency limitations, we divided the NIR range into two sub-bands (700 to 1050nm and 1050 to 1600nm). We implemented adaptive hardware designs for the two sub-bands, and the specially designed reflective coded aperture functions across the full 700–1600 nm range, enabling its reuse and thereby minimizing the complexity of component replacement. The wavelength-segmented design enables seamless broadband acquisition. Additionally, the reflective optical encoding structure reduces the space requirements compared to DD-CASSI, which enhances the system's compactness and flexibility.

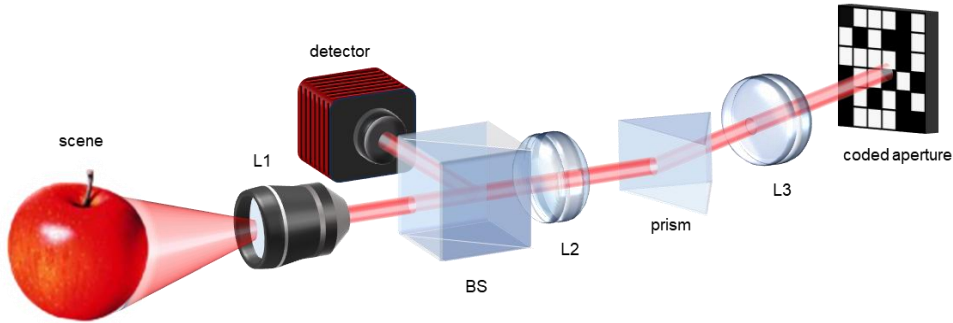

Fig. 1. Schematic of the reflective coded aperture broadband NIR hyperspectral imaging system.

## 2. Methods

### 2.1 Compressed hyperspectral imaging model

Compressed spectral imaging involves encoding the three-dimensional spatial-spectral data cube onto a two-dimensional detector, enabling snapshot acquisition in a single exposure. The imaging models of SD-CASSI, DD-CASSI, and R-CASSI schemes are similar. They differ in their dispersion processes: SD-CASSI involves a single dispersion, DD-CASSI involves dual dispersion, and R-CASSI uses one prism to achieve dual dispersion. R-CASSI combines the strengths of SD-CASSI and DD-CASSI by using a reflective coded aperture.

Fig. 1 illustrates the schematic of the proposed optical system. The light is focused by an objective lens and then dispersed by a prism. A reflective coded aperture encodes the optical information. Subsequently, the reflected light passes through the prism a second time and is reflected into a detector by a beam splitter. The detector captures the two-dimensional compressed measurement, which can be reconstructed into three-dimensional (3D) hyperspectral data according to compressed sensing [21,22].

Hyperspectral data inherently form a 3D cube, where x and y represent spatial coordinates and $\lambda$ denotes the spectral coordinate. Mathematically, let $f(x, y, \lambda)$ denotes the spectral intensity at the spatial coordinates $(x, y)$ for wavelength $\lambda$ and $T(x, y)$ denotes the transmission function of the coded aperture. $\alpha(\lambda)$ is the dispersion function, which expresses the association between spatial displacement and wavelength.

Assuming dispersion occurs along the x-axis, the spectrally dispersed and spatially coded data can be expressed as:

$$f'(x, y, \lambda) = f(x + \alpha(\lambda), y, \lambda) T(x, y), \tag{1}$$

After the second traversal, it becomes:

$$f''(x, y, \lambda) = f(x, y, \lambda) T(x - \alpha(\lambda), y), \tag{2}$$

Since the detector can only record grayscale intensity, the measurement $I(x, y)$ is represented by an integral over wavelength $\lambda$:

$$I(x, y) = \int_{\lambda_{min}}^{\lambda_{max}} f''(x, y, \lambda) \, d\lambda, \tag{3}$$

Let $\mathbb{R}^{N_x \times N_y}$ represents the spatial dimensions of a two-dimensional matrix. The spectral dimension is discretized into $N_\lambda$ bands, forming a set of spectral channels $\{\lambda_1, \lambda_2, \ldots, \lambda_L\}$. The compressed measurement process can be formulated as a linear matrix equation:

$$I = Hf + \omega, \tag{4}$$

Here, $I \in \mathbb{R}^{N_x \times N_y}$ denotes the measurement captured by the detector, $f \in \mathbb{R}^{N_x \times N_y \times N_\lambda}$ denotes the 3D hyperspectral data cube, and $\omega \in \mathbb{R}^{N_x \times N_y}$ denotes the noise term. $H \in \mathbb{R}^{N_x}$ represents the modulation matrix of the system, which is related to the spectrally shifted transmission function $T(x - \alpha(\lambda), y)$.

The ultimate goal is to reconstruct 3D hyperspectral data f from the compressed measurement $I$. A unique solution can be found for this underdetermined equation by optimizing the sparsity-constrained problem:

$$\arg \min_f \frac{1}{2} \|I - Hf\|_2^2 + \tau \Phi(f), \tag{5}$$

Where $\Phi(f)$ is a regularization term and $\tau$ is a weighting factor.

### 2.2 Reconstruction Algorithm

Hyperspectral data reconstruction is solved by solving the optimization problem in Equation (5). Specifically, we employ the two-step iterative shrinkage thresholding (TwIST) algorithm together with a total variation (TV) regularizer [23,24].

The TwIST algorithm is an advanced version of the standard iterative shrinkage thresholding (IST) algorithm. In the TwIST algorithm iteration process, each step relies on the previous two iterations, resulting in a significant acceleration of convergence. The iterative steps for solving the optimization problem in Formula (5) are as follows:

$$f_1 = \Gamma(f_0), \tag{6}$$

$$f_{t+1} = (1-\alpha)f_{t-1} + (\alpha-\beta)f_t + \beta\Gamma(f_t), \tag{7}$$

$$\Gamma(f) = \Psi(f + H^T(I - Hf)), \tag{8}$$

where $\Psi$ is a denoising function. $\alpha$ and $\beta$ are the parameters of the algorithm. By adjusting parameters such as $\Psi$, $\alpha$, $\beta$ and the number of iterations, the algorithm is optimized to achieve the best reconstruction results.

Due to the lack of publicly available NIR hyperspectral datasets, deep learning approaches were not adopted in this work. To demonstrate our system's compatibility with diverse iterative algorithms, we conducted comparative reconstructions using both TwIST and GAP-TV (Generalized Alternating Projection based Total Variation) [25] algorithms on identical experimental data. As shown in Fig. 2, TwIST provided better reconstruction results. Thus, experimental results presented in Chapter 3 were reconstructed using the TwIST algorithm.

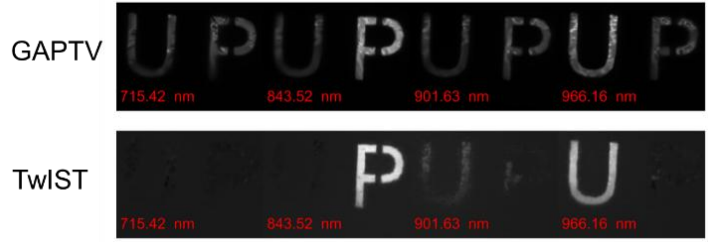

Fig. 2. Comparison of reconstructed results using TwIST and GAP-TV algorithms.

*2.3 Optical Components*

The reflective coded aperture imaging system acquires compressed measurements in the broadband NIR range and reconstructs the 3D hyperspectral data using the TwIST algorithm. To cover a wide range of wavelengths, optical components and structures must be designed to operate efficiently across a broad spectral range, requiring high precision and superior material performance.

Fig. 3 is the physical image of the experimental setup. To acquire compressed measurements with minimal optical inaccuracy, all components of the system are optimized specifically for the NIR region. We use the short-wave infrared focal plane array detector (Cobra2000-CL1280-130VT-00) to capture measurements. The other optical components are coated with spectrally sensitive anti-reflective coatings. The objective lens (L1) is a single-focus lens (M1614-MP2), while L2 and L3 are relay lenses (ACA254-060-B). The prism (PS814-B or CWP14) and the beam splitter (BS014 or MBS1455-C) need to be selectively replaced according to the requirements of sub-bands. By replacing optical components, hyperspectral data of sub-bands can be captured separately. The reflective coded aperture is a binary ('1' and '0') random pattern with a pixel pitch of 10 μm, which is employed to encode the light information. The incident light is modulated by the binary-coded aperture, where transparent regions (denoted as '1') transmit light while opaque regions (denoted as '0') block it, resulting in a patterned intensity distribution. This coded aperture is composed of a silicon dioxide substrate, a silver reflective layer, and a chromium oxide anti-oxidation layer. The reflectivity of Ag is higher than 95% in the range from 700 to 1600 nm. The high reflective property of Ag makes it suitable for reflective coding in the NIR region. With a fabrication error of ±0.2 μm, it remains within the acceptable tolerance for our experiments. To ensure precise alignment, we implemented an innovative cross-shaped reference pattern on the coded aperture, effectively eliminating tilt in the horizontal and vertical axes. The calibrated coded aperture was illuminated with lasers of different central wavelengths to verify that the pixel shift matched the number of spectral channels defined in the simulation. with the calibration error being less than the spectral channel spacing (17 nm).

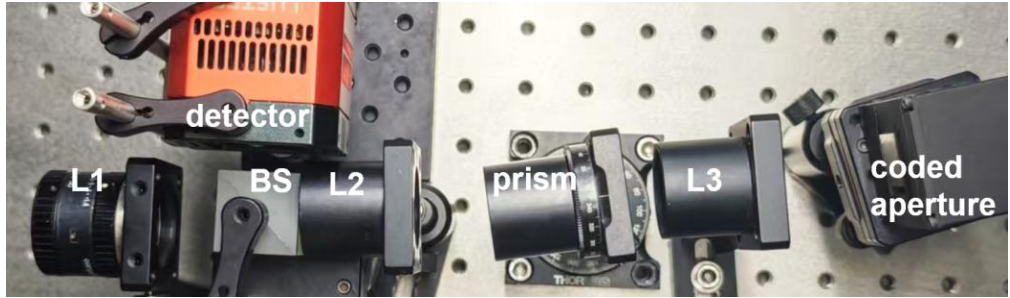

Fig. 3. The physical image of the experimental setup.

The advanced optical structure and component designs facilitate high-precision acquisition in both spectral and spatial domains. In metrics characterizing the spectral and spatial domains, spectral channel spacing is the wavelength interval between adjacent spectral bands, and spatial resolution denotes the smallest distinguishable feature size in images. The proposed system can acquire hyperspectral images across 52 spectral channels in the 700-1600 nm range, with a spectral channel spacing of approximately 17 nm. The spatial resolution primarily depends on the detector and system calibration. In the experimental results of Chapter 3, the highest spatial resolution is 256 × 512.

## 3. Results

In this section, we conduct a series of experiments to validate the spatial and spectral accuracy of the proposed system. The experimental targets included letter-patterned targets and apples representing real-world scenes, while the light sources included lasers and a halogen lamp.

To validate the capability of our system within the 700 to 1600 nm range, we conducted an imaging experiment using letter-patterned targets. The target was illuminated by different light sources: the letters "U" and "P" were illuminated by 850 nm and 950 nm LED lamp beads, while the letters "a" and "b" were illuminated by 1064 nm and 1550 nm lasers. Fig. 4 and Fig. 5 show the letter-patterned targets and experimental results. Fig.4(a) presents the letter-patterned targets illuminated by different light sources. The spatial resolution of the letter combination "up" is 350 × 256, and the spatial resolution of the single letters "a" and "b" is 256 × 256. In Fig.4(b) and Fig.4(c), the peaks of the reconstructed spectral curves closely match the wavelengths of the corresponding illumination sources. Fig.5(a) and Fig.5(b) show that all letter patterns have strong spectral responses and clear spatial profiles near their corresponding wavelengths. These results demonstrate that our system possesses high spectral accuracy and spatial clarity.

Fig. 6 presents the reconstruction results of a real apple and a fake apple captured by our system. The scene was illuminated with a halogen lamp. Fig.6(c) and Fig.6(d) show the reconstructed spectral curves and the ground truth. The spectrum of the real apple exhibits a distinct decline within a spectral range from 900 to 980 nm. In the range exceeding 1350nm, the spectral intensity of the real apple is lower than that of the fake apple. These spectral features, which are consistent with the ground truth measured by a spectrometer, can be used for authenticity identification. The spectral correlation was used to evaluate the reconstruction accuracy. The correlation between the reconstructed spectral curve and the ground truth is 0.936 for real apples and 0.945 for fake apples. Fig.6(a) shows the spectral images of several bands with a spatial resolution of 256 × 512. These images show a significant intensity difference between real and fake apples, in line with the trends in the spectral curves.

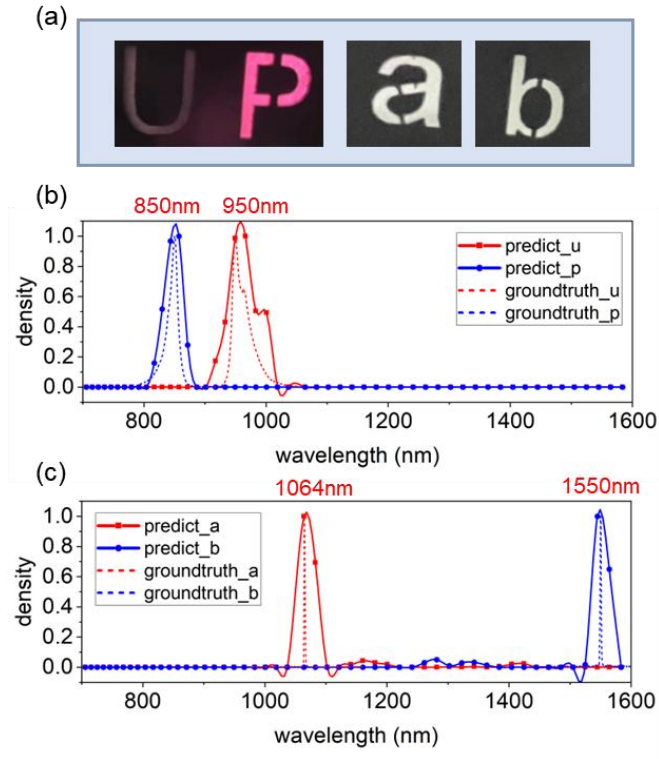

Fig. 4. The letter-patterned targets and experimental results for the letter-patterned targets. "U" is illuminated by an 850 nm LED lamp bead, "P" is illuminated by a 950 nm LED lamp bead, "a" is illuminated by a 1064 nm laser, and "b" is illuminated by a 1550nm laser. (a)RGB reference images. Reconstructed spectral curves of targets and the ground truth. (b) "U" and "P", and (c) "a" and "b". The spectral peaks of the illumination sources are labeled.

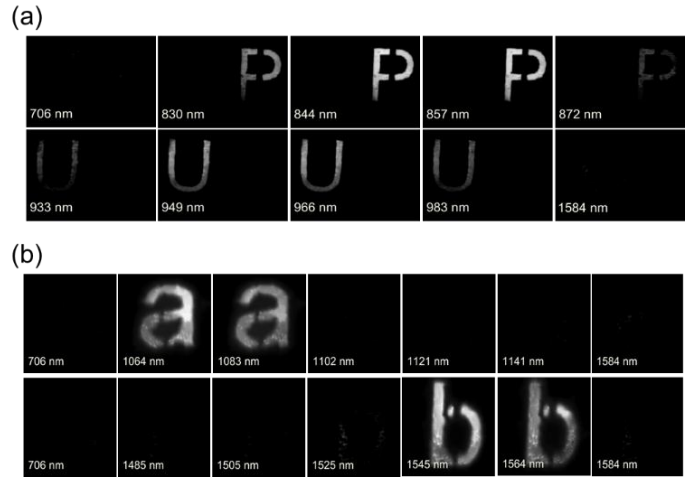

Fig. 5. Reconstructed spectral images near the wavelengths of corresponding illumination sources. All letter-patterns ("P", "U", "a" and "b") have strong spectral responses near their corresponding wavelengths. (a)Spectral images near 850nm and 950nm. (b) Spectral images near 1064nm and 1550nm.

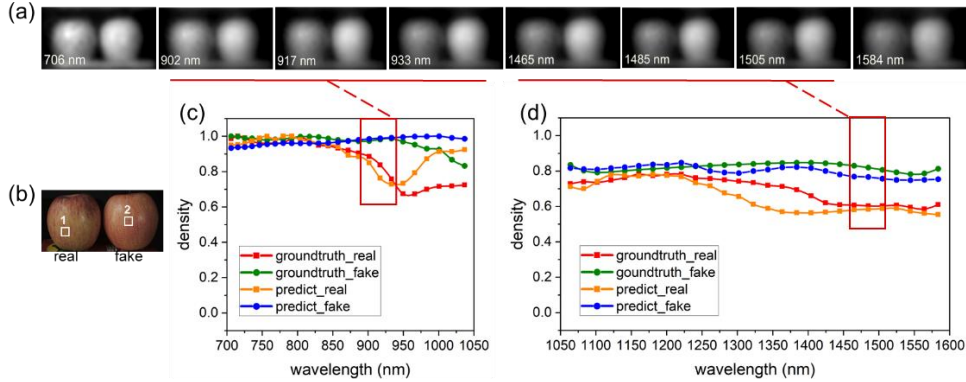

Fig. 6. Experimental results for distinguishing real and fake apples. (a) Some images for specific channels in the 900 to 980 nm spectral range and the range exceeding 1350nm. Spectral images of six bands in the 880 nm to 980 nm range and 1450 nm to 1550 nm range clearly show differences between real and fake apples. (b)RGB reference image. (c) and (d) Reconstructed spectral curves and the ground truth in the range of (c) 700 nm to 1050 nm and (d) 1050 nm to 1550 nm. Two points marked by '1' and '2' in the RGB reference image are selected to plot the spectral curves.

## 4. Conclusion

In summary, a broadband NIR hyperspectral imaging system has been realized, enabling 52-band high-quality NIR hyperspectral imaging with an average spectral channel spacing of 17 nm in the range of 700 to 1600 nm. The wavelength-segmented design expands its applicable spectral range, allowing for the capture of more abundant spectral characteristics. Additionally, the reflective optical structure reduces the system's space requirements, enabling more flexible deployment in practical applications. Experimental results validate the high spectral accuracy and spatial clarity of this approach, highlighting its potential for agricultural applications and anti-counterfeiting technology. Considering that the division into two sub-bands introduces operational complexity, our future work will focus on streamlined sub-band switching mechanisms and automated alignment procedures to simplify implementation without sacrificing performance.

## 5. Back matter

### 5.1 Funding

**Funding.** The Science and Technology Innovation Project for Xiong 'an New Area (NO. 2023XAGG0089) ；National Natural Science Foundation of China (No. 62522502, 62371056) ；Sponsored by Beijing Nova Program；Shenzhen Science and Technology Program (KJZD20230923115202006); the Fund of State Key Laboratory of Information Photonics and Optical Communication BUPT (No. IPOC2025ZZ02); the Fundamental Research Funds for the Central Universities (No. 530424001, 2024ZCJH13).

### 5.2 Disclosures

**Disclosures.** The authors declare no conflicts of interest.